\documentclass[12pt,preprint]{aastex}

\usepackage{rotating}

\slugcomment{Second Submission to AJ}

\shorttitle{Structure of the Cold ISM}
\shortauthors{Holwerda et al.}

\begin{document} 

\title{The Opacity of Spiral Galaxy Disks VIII:\\
Structure of the Cold ISM}
\author{B. W. Holwerda\altaffilmark{1}, B. Draine \altaffilmark{2}, K.D. Gordon\altaffilmark{3}, R. A. Gonz\'{a}lez \altaffilmark{4}, D. Calzetti\altaffilmark{5}, M. Thornley \altaffilmark{6}, B. Buckalew \altaffilmark{7},  Ronald J. Allen\altaffilmark{1} and P. C. van der Kruit \altaffilmark{8}}

\email{holwerda@stsci.edu}

\altaffiltext{1}{Space Telescope Science Institute, Baltimore, MD 21218, USA}
\altaffiltext{2}{Dept. of Astrophysical Sciences, Princeton University, Princeton, NJ 08544-1001, USA}
\altaffiltext{3}{Steward Observatory/Department of Astronomy, University of Arizona, Tucson, AZ 85721 USA }
\altaffiltext{4}{Centro de Radiastronom\'{\i}a y Astrof\'{\i}sica, Universidad Nacional Aut\'{o}noma de M\'{e}xico, 58190 Morelia, Michoac\'{a}n, Mexico}
\altaffiltext{5}{University of Massachusetts, Department of Astronomy, 710 North Pleasant Street, Amherst, MA 01003}
\altaffiltext{6}{Department of Physics \& Astronomy, Bucknell University, Lewisburg, PA  17837}
\altaffiltext{7}{Dept. of Physics, Embry-Riddle University, 3700 Willow Creek Rd, Prescott, AZ  86301, USA}
\altaffiltext{8}{Kapteyn Astronomical Institute, P.O. Box 800, Groningen, the Netherlands}


\begin{abstract}
The quantity of dust in a spiral disk can be estimated using the dust's 
typical emission or the extinction of a known source. In this paper, we 
compare two techniques, one based on emission and one on absorption, 
applied on sections of fourteen disk galaxies. The two measurements reflect, 
respectively the average and apparent optical depth of a disk section. 
Hence, they depend differently on the average number and optical depth of 
ISM structures in the disk.

The small scale geometry of the cold ISM is critical for accurate models of the overall 
energy budget of spiral disks. ISM geometry, relative contributions of different 
stellar populations and dust emissivity are all free parameters in galaxy Spectral 
Energy Distribution (SED) models; they are also sometimes degenerate, depending on 
wavelength coverage. Our aim is to constrain typical ISM geometry.

The apparent optical depth measurement comes from the number of distant galaxies 
seen in HST images through the foreground disk, calibrated with the  ``Synthetic 
Field Method'' (SFM). We discuss what can be learned from the SFM measurement 
alone regarding ISM geometry.  

We measure the IR flux in images from the {\it Spitzer} Infrared Nearby 
Galaxy Survey in the same section of the disk that was covered by HST. 
A physical model of the dust is fit to the SED to estimate the dust surface density, 
mean temperature, and brightness in these disk sections. The surface density is 
subsequently converted into the average optical depth estimate.

The two measurements generally agree and the SED in order model finds a 
mostly cold dust (T $<$ 25 K.). 
The ratios between the measured average and apparent optical depths 
of the disk sections imply optically thin ($\tau_c =0.4$) clouds in these disks. 
Optically thick disks, are likely to have more than a single cloud 
along the line-of-sight.  
\end{abstract}

	
\section{\label{sec:intro}Introduction}

The dust content of a spiral galaxy disk can be mapped either by the 
characteristic dust emission in the far-infrared (FIR) and sub-mm regimes, 
or by using the attenuation of known background sources. Both techniques 
have seen recent significant improvements in accuracy and sensitivity, with 
complementary results shedding light on the dusty interstellar medium 
in spiral disks.

The emission from the interstellar dust in the disks of spiral galaxies has 
been characterized with increasing accuracy by several infrared space 
missions (IRAS, ISO and, recently, {\it Spitzer}), as well as the sub-mm 
observations of SCUBA on the JCMT. The improvements in spatial
resolution and wavelength coverage have led to significant insight into 
the temperature components of the dust in spiral disks, and into the relation 
between dusty clouds and star-formation. 
The FIR emission from spiral galaxies has revealed that the dust can be 
described by two dominant thermal components: warm (25 K $<$ T $<$ 100 K) 
and cold (T $<$ 25 K). Both the warm and cold components can be found in 
spiral arms, and there is a smooth disk of cold dust between these arms. 
Most of the dust mass in the spiral disk is cold \citep[see the review 
articles by][]{Genzel00, Tuffs05rev, Popescu05rev}.

FIR and sub-mm observations of galaxies find indications \citep[e.g.,][]{Trewhella00, Alton98c} 
or direct evidence of cold dust disks extending beyond the stellar disk \citep{Nelson98, Davies99, Popescu03}.
The studies of edge-on spirals by \cite{Radovich01} and \cite{Xilouris99} 
quantified the radial profile as a scalelength of the dust that is 40\% larger 
than that of the starlight. The contribution of the cold ISM (T $<$ 25 K) to 
the overall emission of spiral disks has been difficult to constrain because 
of the degeneracy between dust temperature and mass. Hence, the cold 
ISM's relation with HI and their relative distribution remain unknown.

{\it Spitzer} observations, mostly from the Spitzer Infrared Nearby Galaxy Survey 
\citep[SINGS,][]{sings}, have already contributed greatly to the understanding 
of spiral disks. The relations between the tracers of cold dust (70 and 160 micron 
emission) and star-formation, both obscured (24 micron) and unobscured (UV and 
H$\alpha$ emission) have already been studied with this multi-wavelength survey in 
several canonical galaxies and their substructure: the starburst M51 \citep{Calzetti05,Thornley06}, 
the grand design spiral M81 \citep{Gordon04, PG06}, the rings of NGC 7331 \citep{Regan04} 
and M31 \citep{Gordon06}, the superwind in M82 \citep{Engelbracht06}, and the dwarf 
NGC 55 \citep{Engelbracht04}. \cite{Dale05,Dale07} discuss the SED of all SINGS galaxies 
over all available wavelengths. \cite{Draine07b} find ample evidence for dust in all the 
SINGS galaxies, with the gas-to-dust ratio related to the metallicity. They find no evidence 
for very cold (T $<$ 10K) dust, however.

These studies find ample evidence of cold dust throughout the optical disks of spirals 
but, interestingly, also outside them in various places: on the edge of the optical disk 
\citep{Thornley06,Gordon06}, outside M82's superwind \citep{Engelbracht06} and 
extending beyond the stellar disk \citep{Hinz06}.

Parallel to these investigations of dust emission has gone an observational 
effort to quantify the absorption by dust in spiral disks using known background 
sources. 
\cite{kw92} proposed using occulting galaxy pairs for this purpose. 
Nearby occulting galaxy pairs were initially investigated with ground-based 
data, both images \citep{Andredakis92,Berlind97,kw99a,kw00a} and 
spectra \citep{kw00b}. Subsequently, with the Hubble Space Telescope (HST), 
a more detailed picture of dust in these nearby disks emerged \citep{kw01a,kw01b,Elmegreen01}.
The results of these studies are that extinction is gray\footnote{Gray extinction is 
equal attenuation at all wavelengths: there is no relation between color and 
measured optical depth. A color measurement is dominated by the lines-of-sight with 
the {\it least} extinction, while the independent extinction measure is dominated by those 
with the {\it most} extinction. In cases where many lines-of-sight are mixed, gray extinction 
is mimicked.} when measured over disk sections greater than 100 pc, but resembles the 
Galactic extinction law at smaller scales -- those that can only be resolved with HST. 
Arms are found to be more opaque than the general disk, and some evidence suggests 
that the dust disk is a fractal, similar to the HI disk. 

\cite{Gonzalez98} investigated the use of the calibrated number of distant galaxies seen 
though the foreground disk in HST images. The calibrated counts of distant galaxies have 
been explored further in the previous papers in this series \citep{Gonzalez98,Gonzalez03, 
Holwerda05a, Holwerda05b, Holwerda05c, Holwerda05e,Holwerda05d,Holwerda06sings}. 
Both the occulting galaxy technique and counts of distant objects yield very similar 
opacities for disks and spiral arms \citep[][]{Holwerda05b}.

In recent years, models have been developed to explain the Spectral Energy Distribution 
(SED) of edge-on spirals spanning wavelengths ranging from the UV to the FIR.
\citep[e.g.,][]{Popescu00,Misiriotis01,Popescu00, Misiriotis01,Tuffs04, Boissier04, Dasyra05, 
Tuffs04,Dasyra05, Calzetti05, Dopita06a,Dopita06b,Draine07a,Draine07b}.\\
Three scenarios have been proposed to explain the discrepancy between the 
apparent absorption in UV and optical wavelengths and the emission of dust in the FIR and 
sub-mm regimes:
\begin{itemize}
\item[1.] A young stellar population is embedded in the dense plane of the disk. This is proposed 
by \cite{Popescu00} and corroborated by \cite{Driver07}. The embedded young stars pump the 
FIR emission radiated by the dust plane. 
\item[2.] A strongly clumped dusty medium. The clumping would lead to underestimate the dust 
mass from optical extinction in edge-on systems \citep{Bianchi00b, Witt00, Misiriotis02}; 
the dust mass would also be underestimated by a UV to FIR SED \citep{Bianchi00c}. 
\item[3.] A different emissivity of the cold dust grains --higher than canonical-- in the FIR and 
sub-mm. A change of emissivity has been proposed for denser ISM regions \citep{Alton04, 
Dasyra05} or, alternatively, for the lower density regions of the disk \citep{Bendo06}.
\end{itemize}
In all three of these scenarios, the clumpiness of the dusty ISM is an important factor.

While in some models the large scale structure of the dusty ISM has been somewhat constrained 
\citep[e.g.,][]{Xilouris99,Seth05,Bianchi07,Kamphuis07}, the small-scale geometry (``clumpiness'') 
of the cold ISM remains unknown.
Therefore, an estimate of the prevalent dusty cloud size for spiral disks would provide 
a constraint for the SED models of spiral disks.

Given that SED and extinction techniques are sensitive to different characteristics of the dusty 
clouds in the spiral disk, a comparison between the optical depth derived from these two methods 
has the potential to figure out the structure of the dusty ISM. Here, we compare the $I$-band optical depths 
for a section of the spiral disk, one derived from a SED model of the {\it Spitzer} fluxes (``average''), 
and one determined from the number of distant galaxies found in an HST image (``apparent'').
The term ``average'' refers to what the optical depth proportional to the dust mass,  
uniformly distributed over the disk section. The term ``apparent'' means the effective optical depth
of the disk section for a background uniform light source \citep[original definitions from][]{Natta84}. 
The term ``opacity'' is used throughout our previous papers for the apparent optical depth 
measured over a section of the disk for its whole height.


This paper is structured as follows: \S \ref{sec:data} discusses the {\it Spitzer} 
and HST data used. In \S \ref{sec:analysis}, the two different methods to derive 
optical depths are presented. We discuss the relation between dust geometry 
and galaxy counts in \S \ref{sec:csize}. In \S \ref{sec:results}, we present the 
derived optical depths; \S \ref{sec:model} presents a simple geometric model 
to interpret the results, and \S \ref{sec:concl} lists our conclusions and future work.

\section{\label{sec:data}Data}

The data for this paper come from two archives, the HST archive and the fourth 
data release \citep{dr4} of the {\it Spitzer Infrared Nearby Galaxy Survey} 
\citep[SINGS,][]{sings}\footnote{http://sings.stsci.edu}. There is an overlap of fourteen galaxies between the 
SINGS sample and that of \cite{Holwerda05b}. Two of these, NGC 3621 and 
NGC 5194, have two WFPC2 exposures analyzed in \cite{Holwerda05b}. The 
HST/WFPC2 data reduction is described in \cite{Holwerda05a}. The reasoning 
behind the HST sample selection from the archive is explained in \S \ref{ssec:sfm}.

The overlap between HST and {\it Spitzer} data is illustrated in Figure \ref{fig:overlap}, 
with the WFPC2 footprint projected on the 24 micron {\it Spitzer} images. Only the solid 
angle covered by the WF chips is used for further analysis (the PC chip is excluded).

The Infrared Array Camera (IRAC) mosaic is made with the custom {\it SINGSdither} 
script, by M. Regan, that combines the scan images into a single mosaic using the 
``drizzle'' algorithm \citep[Fourth Data Release Notes,][]{dr4}. 
The Multiband Imaging Photometer for {\it Spitzer} (MIPS) data products are calibrated, 
sky--subtracted mosaics in all three bands, reduced as described in \cite{Gordon05, 
Bendo06, dr4}. The basic instrument parameters, pixel scale, and adopted PSF FWHM for the 
seven main {\it Spitzer} imaging modes are summarized in Table \ref{table:data}. 

\section{\label{sec:analysis}Analysis}

Two parallel estimates of the optical depth of disks are used in this paper: 
first, the apparent optical depth of the spiral disks is determined from the number of distant galaxies 
identified in the HST/WFPC2 images, calibrated with the ``Synthetic Field Method'' (SFM). 
Secondly, the optical depth of the same section of the spiral disks is derived 
from the dust surface density, which is a result of the SED model fit to 
the {\it Spitzer} fluxes using the model from \cite{Draine07a}.

\subsection{\label{ssec:sfm}Galaxy counts: ``Synthetic Field Method''}

In principle, the number of distant galaxies seen through a spiral disk is a 
function of the dust extinction, as well as the crowding and confusion by the 
foreground disk. Initial applications of the number of distant galaxies as 
an extinction tracer were on the Magellanic Clouds \citep{Shapley51, 
Wesselink61b,Hodge74, MacGillivray75}, but they lacked accuracy. 
The ``Synthetic Field Method'' (SFM) was developed by 
\cite{Gonzalez98} to correct an extinction measurement based on the 
number of distant galaxies in an HST image, for the effects of crowding 
and confusion by the foreground spiral disk. 

The ``Synthetic Field Method'' follows a series of steps. First, the 
number of distant galaxies in an HST science field is determined. Second, a series 
of simulated (``synthetic'') fields are made. In each of these fields, a typical 
background (e.g., the Hubble Deep Field) is first dimmed by a gray screen and 
added to the science field. Third, the added distant objects are identified in 
these synthetic fields. The fourth step is to measure the relationship between the 
number of these identified synthetic distant galaxies and background 
dimming. From this relation and the original number of actual distant 
galaxies found in the science field, an average opacity can be found.
It is important to remake the synthetic fields for each science field because the 
crowding and confusion issues are unique in each case.

An additional uncertainty in the resulting average extinction measurement 
is the cosmic variance in the intrinsic number of distant galaxies behind the 
foreground disk. The uncertainty due to cosmic variance can be estimated from 
the two-point correlation function of distant galaxies and folded into the Poissonian 
error. The cosmic variance uncertainty is of the same order as the Poisson statistical 
error for small numbers \citep[For a complete discussion on the uncertainties of the 
SFM, see][]{Holwerda05a}. Therefore, single-field SFM measurements remain 
uncertain, but a meaningful conclusion can be drawn from a combined set of 
science fields.

We have applied this method successfully on archival WFPC2 data. 
\cite{Holwerda05b} present the average radial opacity profile of spiral disks and the 
effect of spiral arms. The spiral arms are more opaque and show a strong 
radial dependence, while the more transparent disk shows a flat profile.
\cite{Holwerda05c} compare HI radial profiles to the opacity ones, and conclude 
that no good relation between disk opacity and HI surface density radial profiles can 
be found. However, \cite{Holwerda05c} find that the sub--mm profile from \cite{Meijerink05} 
generally agrees with their opacity measurements of M51. 
\cite{Holwerda05d} compare the relation between surface brightness and disk 
opacity; this relation is strong in the spiral arms, but weak in the rest of the disk.

\cite{Gonzalez03} predicted, based on simulated data, that the optimum distance for 
the application of the SFM with current HST instruments is approximately that of Virgo. 
The identification of background galaxies suffers in closer disks, as the stellar disk 
becomes more resolved, compounding confusion. This optimum distance, combined 
with the availability of deep HST/WFPC2 images from the Cepheid Extragalactic Distance 
Scale Key Project, resulted in the sample presented in \cite{Holwerda05b}. \cite{Holwerda05e} 
confirmed the results from \cite{Gonzalez03} using this sample, with foreground 
disks spanning distances between 3.5 and 35 Mpc. A selection effect of the Key Project 
is that the majority of the HST science fields is concentrated on spiral arms and exclude 
the centers of the galaxies.

Here, we present average extinction values for the whole WFPC2 field--of--view, 
minimizing the uncertainties to the extent possible. 
In our initial papers, we did not apply an inclination correction to the optical depths because
the correction depends strongly on the dust geometry \citep[see the discussion in ][]{Holwerda05b}. 
However, in this paper we assume a simple dust model in \S \ref{sec:model}.
The appropriate inclination correction --$\times ~ cos(i)$-- has been applied to the points 
in Figures \ref{fig:fit} and \ref{fig:n_val}, and in Table \ref{table:A}. The uncertainties in the 
tables and figures reflect the 1-sigma confidence levels produced by the combination of 
the Poisson error and the cosmic variance of background galaxies.

\subsection{SED optical depth estimate}

\begin{figure}[htbp]
\begin{center}
\includegraphics[width=\textwidth]{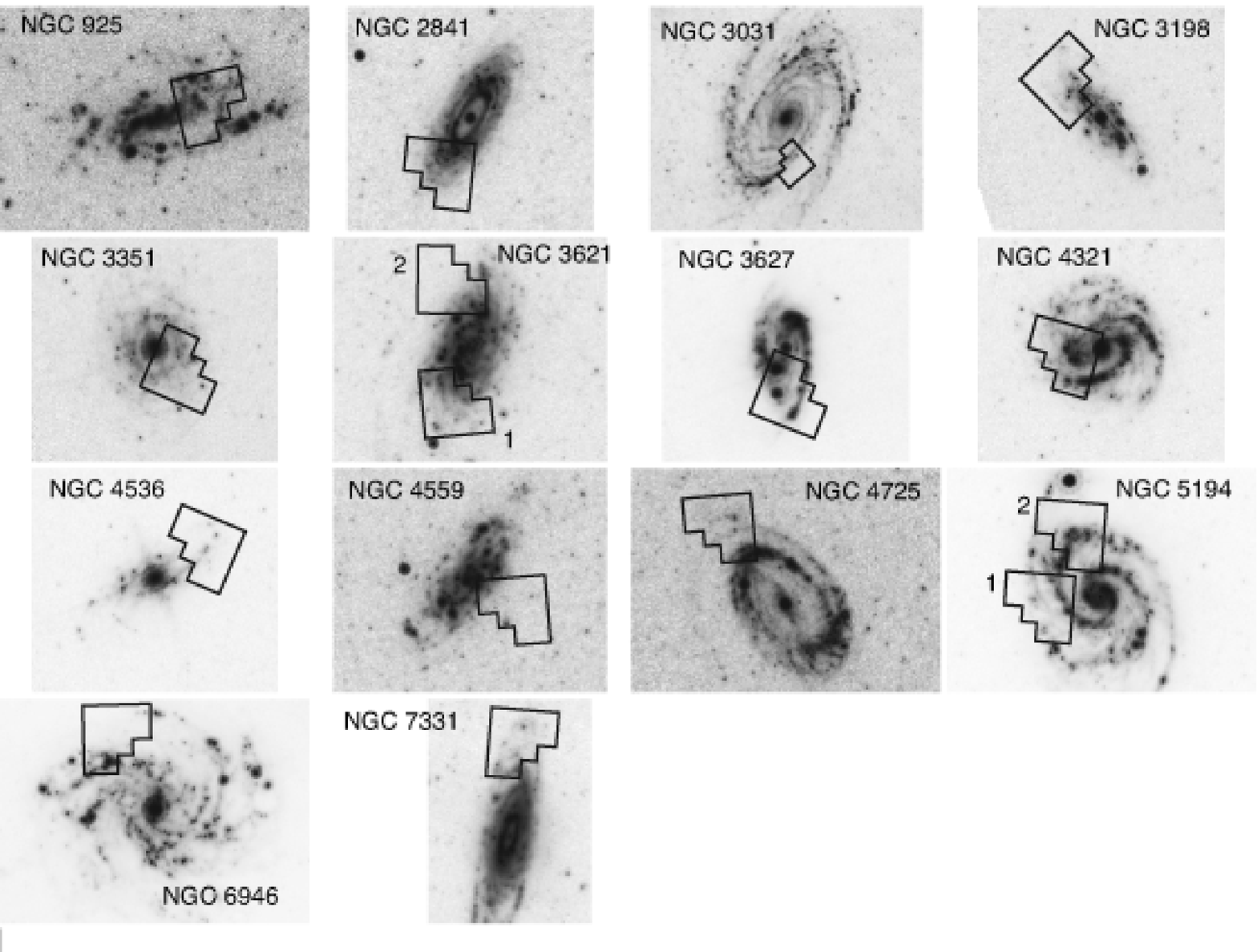}
\caption{The footprints of the WPFC2 camera on board HST, on the 24 micron images 
from the MIPS detector onboard {\it Spitzer}. Most of the HST images do not include the center, and are pointed on a spiral arm. NGC 3621 and NGC 5195 (M51) have two separate WFPC2 fields associated with them. The PC chip, i.e, the small chip in the nook of the ``L'' of the three WF chips, is not used for the SFM analysis, nor as part of our {\it Spitzer} aperture. }
\label{fig:overlap}
\end{center}
\end{figure}

The average disk optical depth is derived from the {\it Spitzer} observations.
First, the surface brightnesses within the WF chips' footprint are measured\footnote{
The PC part of the WFPC2 array is not used in either analysis.} (see Figure 
\ref{fig:overlap}). Second, these are converted into a dust surface density using an 
SED model. Third, this surface density is translated into an $I$-band optical depth.

All the IRAC and MIPS data are convolved to the poorest resolution of the 160 
micron observations (see Table \ref{table:data}), and the pixel scale is set to 9". 
This is done with the  {\it gauss, wcsmap} and {\it geotran} tasks, under IRAF.
Subsequently, the overall flux is measured in the WFPC2 field of view (Figure 
\ref{fig:overlap}). Because the 
L-shaped aperture is a highly unusual one, the aperture correction remains 
uncertain, but not negligible, since the FWHM at 160 micron ($40\farcs0$) 
is of the order of the aperture diameter ($3 \times 1\farcm3 \times 1\farcm3$ in an L-shape).
Published aperture corrections for the IRAC instruments \citep{Hora04} 
overestimate the correction for extended objects \citep{IRACaper}. Here we use 
the aperture corrections for extended sources from \cite{IRACaper} for IRAC 
fluxes,\footnote{http://spider.ipac.caltech.edu/staff/jarrett/irac/calibration/index.html} 
and from \cite{MIPSaper} for the MIPS fluxes\footnote{http://ssc.spitzer.caltech.edu/mips/apercorr/} 
(see also Table \ref{table:data}).
Table \ref{table:flux} gives the average surface brightnesses for the seven 
{\it Spitzer} channels in the field-of-view of the three WF chips of the WFPC2
array. The uncertainties are derived from the variance in the sky. 
Generally, the surface brightnesses agree with the results presented by 
\cite{Dale05} for the entire disks.

\begin{figure}[htbp]
\begin{center}
\includegraphics[width=\textwidth]{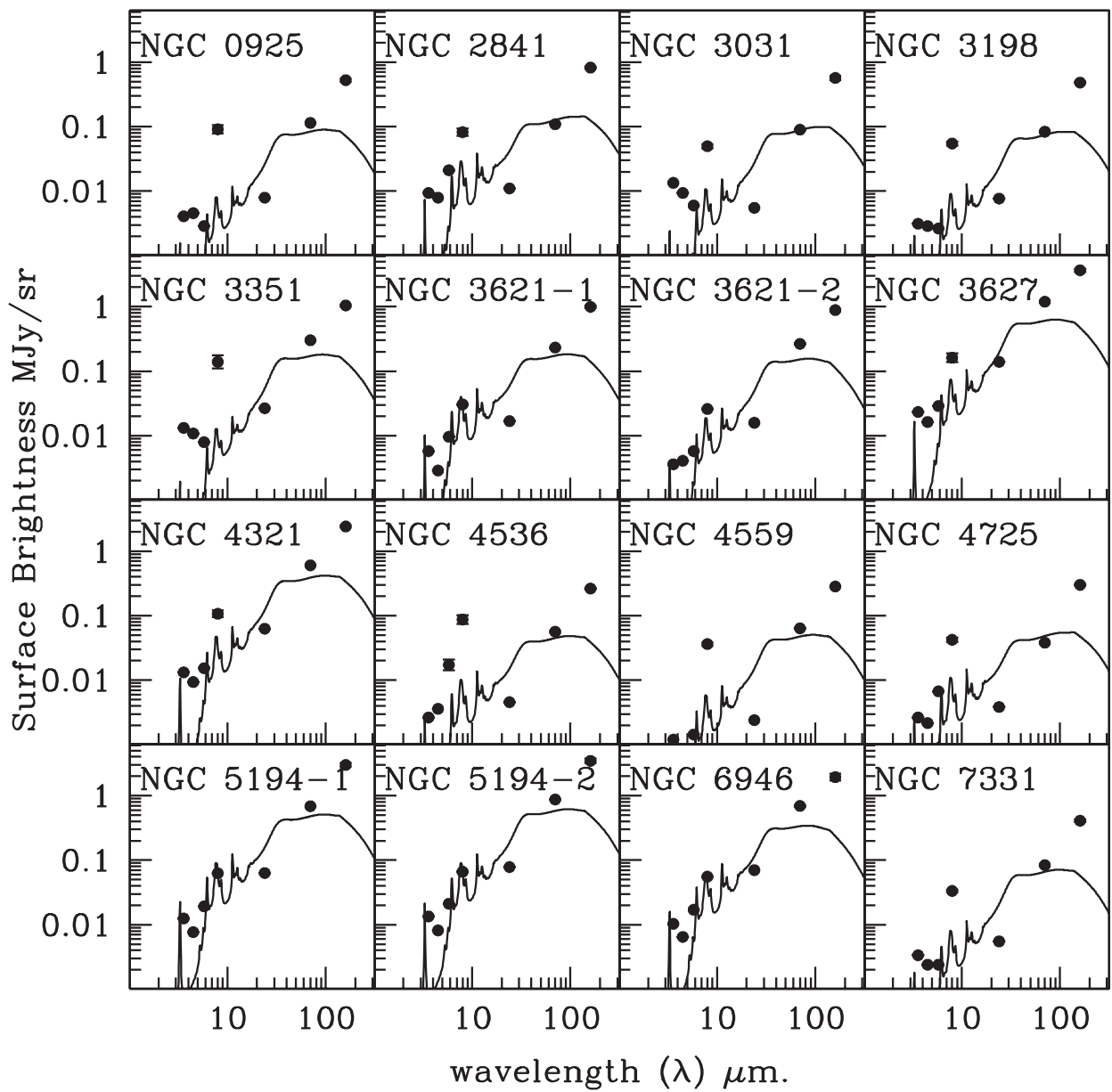}
\caption{The Spectral Energy Distribution in the 7 {\it Spitzer} bands (IRAC/MIPS), for each of the WFPC2 apertures in Figure \ref{fig:overlap}. The best fitting model from \cite{Draine07a} is shown. The relevant parameters for each fit are in Table \ref{table:model}.  
}
\label{fig:spec}
\end{center}
\end{figure}

The second step is to convert these surface brightnesses to a dust surface density. 
Initially, we fitted only the MIPS fluxes with two blackbodies and derived surface densities 
from these \citep{Holwerda06sings}. However, a more rigorous treatment of the 
IR fluxes can be done with a SED model, such as the one presented in \cite{Li01}. 
This model uses the physics of grain heating and reradiation, and a model distribution 
of grain sizes and types. The updated version from \cite{Draine07a} has been fit to the data, 
and the results are shown in Figure \ref{fig:spec}. The derived stellar and dust surface 
brightnesses, dust surface density, and mean temperature are presented in Table 
\ref{table:model}. Dust surface densities are between 0.1 and $1.4 \times 10^6 M_{\odot} ~ kpc^{-2} $, 
with mean temperatures between 14.6 and 17.8 K. 

The mean dust temperatures are obtained from the mean radiation scaling, $\bar{U}$,
\footnote{The \cite{Draine07a} model uses a distribution of scaling values ($U$) of the local Interstellar Radiation Field 
to calculate the irradiation that the grains see. $\bar{U}$ is the average of this scaling distribution.} 
in the \cite{Draine07a} model ($T_2$ in their equation 18). The model uses a distribution of 
temperatures and grain sizes. Therefore the mean temperature is an indication of the 
thermal equilibrium point of the bulk of the dust. Most of the dust is cold ($\rm T < 25 ~ K$). 

These results are an obvious improvement over a simple single-temperature fit, but the 
model parameters in Table \ref{table:model} are still not fully constrained; disagreement 
between data and model at the PAH peak at 8 micron could be an effect of metallicity or 
the presence of a bright HII region. The 70 and 160 $\mu$ m fluxes hint at colder or more dust 
in the disk (Figure \ref{fig:spec}). There are three caveats to the fits: (1) a lack of sub--mm data, 
(2) the single color ISRF used and (3) averaging over different types of emission regions, i.e., 
HII regions and the general disk.

Additional sub-mm data of comparable quality, needed to better constrain the model and 
especially the cold dust emission, will not be available until SCUBA2 starts operations on 
the JCMT and the launch of {\it Herschel}.\footnote{{\it Herschel} data will be especially 
valuable will be especially valuable as it will not suffer from night sky structure, which is of 
similar angular size as these disks.} At present, comparable-quality sub-mm 
maps are available only for NGC 5194 \citep{Meijerink05} and NGC 7331 \citep{Regan04}. 
\cite{Draine07b} discuss SED models with and without sub-mm data. 
%

The second caveat in the derivation of dust mass is the assumption of a constant color for 
the interstellar radiation field (ISRF) illuminating the emitting dust. In reality, dust grains 
deeper in a dust structure will encounter radiation field that is not only dimmed but also 
reddened and, hence, will contribute less flux to the FIR emission. Locally, the ISRF will 
also depend on the age of the nearby stellar population. Due to the reddening of the ISRF 
deeper in the cloud, dense clouds could contain more dust mass in their centers than 
inferred from just the FIR emission. Additional sub-mm observations will help resolve 
this uncertainty in the SED optical depth. \cite{Draine07b} discuss fits of the model to 
the total fluxes of the SINGS galaxies with and without additional sub-mm data. They find 
that the FIR estimate underestimates of 1.5 times the dust mass more (5 out of 17 cases, 
notably NGC 3627 and 7331 of our sample) than overestimates it (only M51 of the 17). 

The third caveat is that the model values are an average over many different types of ISM 
regions, each with a different heating mechanism, dust structure and composition (e.g., 
photo-dissociation regions, cirrus, and star-forming regions in spiral arms). 
The \cite{Draine07a} model's assumptions hold better for some regions than for others, 
but we use the results as ``typical'' for these disks. The relative contribution of PAH 
emission to the SED is a function of ISM geometry as well as irradiation and composition 
\citep[e.g., ][]{Silva98,Piovan06}. Together with a better-constrained FIR/sub-mm SED 
one could constrain ISM geometry solely from the relative contributions to the SED. 
\cite{Draine07b} discuss the application of the \cite{Draine07a} model to whole disks 
of the SINGS galaxies.

The dust surface density is translated into an average optical depth with the 
absorption cross section per unit dust mass, $\kappa_{abs}(\lambda)$, and 
grain albedo from \cite{Draine03} for the Johnson $I$-band (865.5 \AA.): 
$\tau_m = \kappa_{abs} \times (1-Albedo) \times M_{dust}/area$.
These optical depth values are presented in Table \ref{table:A}.

\section{\label{sec:csize} Cloud size and the SFM}

It continues to be difficult to constrain dusty cloud geometry from models of either  
extinction or emission. In this section, we review what can be learned, solely 
from the SFM measurements, about the geometry of the extincting medium.

In \cite{Holwerda05b}, two indications that the dust disk is clumpy are identified: 
(1) the average color of the distant galaxies is independent of the disk opacity implied 
by their number, and (2) the measurement of disk opacity is independent of inclination. 
The lack of a relation between the average opacity of the disk and the average color 
of the detected galaxies can be explained by two scenarios: (1) some of the 
background galaxies are completely blocked by large clouds, and some are not. 
The color measurement is done on background objects that do not suffer from 
extinction and, hence, reddening. Alternatively, (2) {\it all} background galaxies are 
dimmed by clouds smaller than the projected distant galaxies. Consequently, 
some of the distant galaxies are dimmed enough to drop below the detection threshold. 
Any detected galaxy's color is, however, measured from mostly unreddened 
flux.\footnote{The effect of cloud geometry on the luminosity function of the detected 
distant galaxies is, in principle, another possible way to distinguish between cloud 
geometries, but this is complicated by the fact that the detection limit of distant galaxies is dominated by 
the field properties (brightness and granularity), rather than by the dimming by the dust. 
See also the discussions in \cite{Holwerda05e}, and Appendix B in \cite{Holwerda05}.}  
However, we note that a relation between the reddening and derived extinction 
from the distant galaxies is difficult to detect because (1) the spread in colors of distant 
galaxies is substantial and (2) color is measured from the detected galaxies 
--automatically the least dimmed-- whereas opacity is measured form the {\it missing} galaxies. 

The inclination effect on the apparent optical depth from the number of 
distant galaxies is minimal \citep[see ][, \S 5.1 and Figure 3]{Holwerda05b}. 
Assuming a thin layer of optically thick clouds, the apparent optical depth of the disk, 
measured from the number of distant galaxies, is dominated by the apparent filling 
factor of clouds. The projected filling factor does not change much with inclination: 
a flat cloud covering 40\% of a certain disk section still covers 40\% of the inclined section. 
Only when the height of the cloud becomes important --when the inclination is closer to 
edge-on--, does the apparent filling factor change. This explanation for the lack of an 
inclination effect in the opacity profiles does not depend on the size of the clouds. 
It could be a single, large, cloud or many small ones in the plane of the disk.
However, the optical depth values in \cite{Holwerda05b} are from different sections of 
the disks --although generally centered on a spiral arm-- and the effect of small inclination 
differences could well have been masked by comparing different regions in the disks. 
In \cite{Holwerda05b} we did not apply an inclination correction because it depends on the 
assumed dust geometry. In this paper we do assume a dust geometry and hence make 
an inclination correction (\S \ref{sec:model}).

The simulations in the SFM assume a gray screen, an uniform unclumped dust layer with 
opacity equal in the V and I bands. The SFM opacity measurements in this paper are based 
on such simulations. In \cite{Holwerda05}, we ran a series of simulations on 
NGC 1365\footnote{NGC 1365 is one of the more distant galaxies in the \cite{Holwerda05b} 
sample, and we ran these simulations as a validation of the gray screen synthetic fields.} 
to characterize the effect of average cloud cross-section on the number of distant galaxies 
observable through a disk. 

Figure \ref{fig:sfmsize} shows the effect of cloud size, expressed in pixels, on the simulated 
relation between average opacity and number of distant objects. In each simulation, we fix 
a cloud size and vary their number to increase disk opacity. An ensemble of unresolved 
clouds is effectively the gray screen. For clouds resolved with HST, more than 2 
pixels\footnote{Pixel scale is $0\farcs05$ in our data, after the drizzle reduction.}, the relation 
between opacity (cloud filling factor) and number of distant objects becomes much shallower. 
The same number of distant galaxies observed would then imply a much {\it higher} opacity 
of the disk. Because the SFM (calibrated with a gray screen) generally agrees well with 
measurements from overlapping galaxies \citep{Holwerda05b}, it seems 
unlikely that the disk's opacity is predominantly due to large, resolved clouds. 
A pixel of $0\farcs05$ at the distance of NGC 1365 (18 Mpc) is 4 pc in linear size. It is therefore 
implied that the structure of the ISM responsible for the disk opacity measured with the 
SFM varies in optical depth on scales of $\sim$10 pc or less.

From the SFM measurements alone, the cloud geometry is impossible to determine. Only when 
additional information is used --e.g., the general agreement with the occulting galaxy technique-- 
it favors small (unresolved) scales for the clouds. Therefore, to constrain cloud geometry, 
information from two different techniques needs to be combined.

\section{\label{sec:results}Optical depths}

Table \ref{table:A} presents the optical depth estimates from the Synthetic Field Method 
and the SED model \citep{Draine03, Draine07a}, for the WFPC2 field-of-view. The optical depths range 
between 0.1 and 3.5 magnitudes in the $I$-band. The measurements are for different 
parts of different spiral disks (Figure \ref{fig:overlap}), explaining in part the range in values.

The optical depth estimates presented here may appear high for the Johnson $I$-band, 
compared with other extinction estimates (e.g., those from inclination effects or 
reddening), but these are (1) for the entire height of the disk, and (2) generally centered 
on a spiral arm. Typical extinction values in the $I$-band are several tenths of a magnitude for a dust screen 
in front of the stellar spiral disk \cite[e.g.,][]{Martin06}. Two of the derived SFM opacities 
are negative, possibly the effect of an overdensity of distant galaxies behind the target galaxy. 
The discrepancy in NGC5194-1 may be due to misidentification of  background galaxies, 
as they are difficult to identify in this field.

Figures \ref{fig:fit} and \ref{fig:n_val} show the values of disk opacity by both methods, over the same 
section of the disk. Both methods generally agree within the uncertainties of the measurements. 
The agreement is better than our initial estimate from a blackbody fit to the MIPS 
fluxes in \cite{Holwerda06sings}. The general agreement and the mean temperature of the 
dust (Table \ref{table:model}) imply that most of a disk's opacity is due to the cold dust 
in the disk.

\section{\label{sec:model}Model of cloud geometry}

The relation between the apparent and average $I$-band optical depth measurements 
--the first from the number of distant galaxies in HST images and the second derived from the 
{\it Spitzer} SED-- could reveal the nature of the prevalent structure in the ISM.

The {\it a priori} assumptions are that (1)  all dust structure is transparent to the FIR emission 
from which the dust surface density is estimated in the SED model, and (2)  the {\it entire} 
volume of the cloud emits in the FIR. We adopt {\it Model C} from \cite{Natta84}, 
in which a randomly distributed series of clumps covers the area. These authors define 
two optical depths: (1) the typical optical depth ($\tau_m$), i.e., the average optical depth that is 
proportional to the dust mass; and (2) the apparent optical depth ($\tilde{\tau}$), or the optical 
depth if a uniform layer would cover the area.

Our two measurements of optical depth --SED and SFM-- correspond to these two optical 
depths, average and apparent. An optical depth based on the SED depends on the dust mass 
within the area, and hence corresponds to $\tau_m$. The SFM optical depth is the apparent optical 
depth averaged over the area, and hence $\tilde{\tau}$. 
These two optical depths need not be the same, and their relation is an indication of 
how clumped the medium is. 

Let us assume a number of small dust structures with a height ($h$), an average grain cross-section ($\sigma$), and a grain emissivity ($Q$). The grain number density in the clouds is denoted by $n_d$, and 
the average number of clouds in a line-of-sight is $n$. We assume all 
clumps have the same optical depth $\tau_c$:

\begin{equation}
\label{eq:tc}
\tau_c = h ~ n_d ~ \sigma ~ Q.
\end{equation}

\noindent The average optical depth ($\tau_m$) is then:

\begin{equation}
\label{eq:tm}
\tau_m = n \times \tau_c,
\end{equation}

\noindent and the apparent optical depth can be derived if one assumes a Gaussian distribution 
of the number of clouds along the possible lines-of-sight and sum the contributions of all clouds 
\citep[see equations 15 and 17-19 in ][]{Natta84}:

\begin{equation}
\label{eq:tap}
\tilde{\tau} = n \times (1- e^{-\tau_c}),
\end{equation}

\noindent with asymptotic values for both optically thin and thick clouds:

\begin{equation}
\label{eq:optthin}
\tau_c \rightarrow 0, ~ \Rightarrow ~  \tau_m \rightarrow 0, ~  \tilde{\tau} \rightarrow \tau_m, 
\end{equation}

\begin{equation}
\label{eq:optthick}
\tau_c \rightarrow \infty, ~ \Rightarrow ~  \tau_m \rightarrow \infty,  ~ \tilde{\tau} \rightarrow n.
\end{equation}

\noindent We note that the inferred dust mass in the disk ($M_{dust}$) is proportional to the number of clouds ($n$) and the cloud optical depth ($\tau_c$).
The ratio of the apparent over the average optical depth is:

\begin{equation}
\label{eq:rat}
{\tilde{\tau} \over \tau_m} = {1-e^{-\tau_c} \over \tau_c},
\end{equation}

\noindent leaving only the optical depth of the clouds ($\tau_c$), and hence the grain density ($n_d$) and the cloud size ($h$) as the variables. Our fit to the relation between SFM and SED optical depth estimates only has $\tau_c$ as the variable (Figure \ref{fig:fit}). 

We can now use equation \ref{eq:rat} to derive $\tau_c$ from a fit to the optical depths from SED 
($\tau_m$), and from the number of distant galaxies ($\tilde{\tau}$). 
We want to answer three questions. 
Are the clouds in the disks typically optically thin or thick? 
If all disks are equal, what is the implied cloud optical depth? 
How many clouds lie typically along a given line-of-sight?

\subsection{\label{ssec:tc}Optically thick or thin clouds?}

\noindent Optically thin clouds ($\tau_c << 1$) result in a ratio of optical depths close to unity:

\begin{equation}
{\tilde{\tau} \over \tau_m} = {1-e^{-\tau_c} \over \tau_c} \approx {1 - (1-\tau_c) \over \tau_c} = 1;
\end{equation}

\noindent optically thick clouds ($\tau_c >> 1$) result in a ratio of:

\begin{equation}
{\tilde{\tau} \over \tau_m} = {1-e^{-\tau_c} \over \tau_c} < 1.
\end{equation}

Figures \ref{fig:fit} and \ref{fig:n_val} show how a majority of the data exhibit a ratio of order unity. 
The cold dust mass in the SED model could be better constrained with additional sub-mm information \citep{Draine07b}. 
However, the ratios are, within the errors, consistent with optically thin clouds in most of the disks.

\subsection{\label{ssec:h}Cloud size}

Figure \ref{fig:fit} shows the fit to the ratios of apparent to  average optical depths, with $\tau_c$ 
as the single fit parameter, as per equation \ref{eq:rat}. For simplicity, we assume here that all 
disks are made up of similar clouds, and that there is no difference between arm and disk regions. 
The negative SFM measurements are excluded from the fit. The best fit is for a cloud optical depth 
of $\tau_c = $0.4 with, on average, 2.6 clouds along the line-of-sight (if the two negative points are 
included, the values change, respectively, to $\tau_c$ = 0.56 and n = 1.9). The inferred value 
of $\tau_c$ is likely to be a mean between higher values in spiral arms and much lower values 
in the disk. Optically thick disks have more clouds along the line of sight, while optically thin disks 
harbor a single cloud. 

Reasonable values for the parameters in equation \ref{eq:tc} are:
$\rm n_d \approx 5 ~ \times 10^{-3} ~ grains ~ m^{-3}$, 
$\rm \sigma = 0.03 ~ \mu m^2$ and $Q= {3 \over 1300} \left({125 \over 160}\right)^\beta = 1.5 ~ 
\times 10^{-3}$ with $\beta = 2$ \citep{Hildebrand83}\footnote{There is substantial 
discussion in the literature about the value of $\beta$ \citep[e.g.,][]{Bendo03}. 
$\beta$ was fixed at 2 in the \cite{Draine07a} models.}. 
The value of 0.4 for $\tau_c$ implies a cloud height, $h$, of $\sim$ 60 pc! 
Much larger clouds could be resolved in extinction maps of these disks based on 
stellar reddening. In the case of NGC 3627, NGC 5194, NGC 6946 and NGC 7331, 
there is a clear spiral arm in the reddening map \citep[][]{Martin06, Holwerda07b}; 
the other reddening maps are smooth or do not extend out to cover the whole WFPC2 pointing. 

A more likely scenario is that there is an inverse relation between the cloud 
density $n_d$ and scale $h$. Such a relation can be seen in giant molecular clouds (GMC) 
of our own Galaxy \citep[e.g, ][]{Solomon87}. In this case, the single optically thin 
cloud can be replaced by smaller optically thick ones. We note that the largest GMCs 
are of the order of 60 pc.


The value of ~60 pc clouds appears in contradiction to the implied size in the SFM calibration 
($\sim$10 pc). However, the typical cloud size can be 60 pc and still the disk opacity can change 
over smaller scales, if several partially overlapping clouds are seen in projection. The inverse 
relation between scale and density would also help make the two scales compatible.

  
Our data are consistent with optically thin clouds ($\tau_c =0.4$) and 
their average opacity value implies a typical cloud size that is unresolved with {\it Spitzer} 
in our galaxies. 

\subsection{\label{ssec:n}Cloud Numbers}

The above fit to the relation between $\tilde{\tau}$ and $\tau_m$ indicates that, on average, more 
than one cloud is needed along the line-of-sight in most disks ($\bar{n} = 2.6$). However, the 
assumption was that  $\tau_c$ had a single value for all the disks. In \S \ref{ssec:tc}, we argued 
that the ratio between $\tilde{\tau}$ and $\tau_m$ implied the $\tau_c$ is optically thin. It logically 
follows that optically thick disks must have more than a single cloud along the line-of-sight. 

Figure \ref{fig:n_val} illustrates the effect of number of clouds along the line-of-sight ($n$) on the 
relation between $\tilde{\tau}$ and $\tau_m$, when $\tau_c$ is freely increased from 0 to $\tau_m/n$ 
(the lines have been drawn according to equations \ref{eq:tm} and \ref{eq:tap}). 
We note that all disks are consistent with many clouds along the 
line-of-sight --including the optically thin ones, also because they lie in the optically thin cloud 
(${\tilde{\tau} / \tau_m} \sim 1$, $\tau_c < 1$) regime.

\subsection{Potential Improvements}

There are many refinements to be made to the simple model presented here. 
Some improvements for future comparisons between these two measurements 
of optical depth are: 
(1) a distribution of cloud sizes for both the SFM calibration, as well as in the model 
explaining the ratio between SFM and SED optical depths. A model distribution can 
be taken from observations of GMCs in our own Galaxy and nearby ones \citep{Heyer01, 
Rosolowsky05}. The cross-section distribution could be used in SFM measurements in the future. 
%
A cross-section distribution can only be applied, if  the foreground disk is at a 
single, fixed distance; only one counts though a single foreground galaxy are used. 
There are three face-on spirals with enough solid angle in HST imaging as well as 
additional {\it Spitzer} data: M51, M81 and M101.
(2) The SED model can be much better constrained with additional sub-mm 
observations, the opportunities for which will expand dramatically in the near 
future ({\it SCUBA2} on the JCMT and {\it Herschel} satellite).
(3) The effects of grand design spiral arms and galactic radius could be identified 
in a single disk; the comparison SFM and SED is not made for different sections of the disks combined.
(4) A future SED model can take into account the reddening of the interstellar radiation field, as it penetrates the ISM.
This would require a comprehensive treatment of the ISM structure in addition to its temperature, composition and irradiation.

\section{\label{sec:concl}Conclusions}

\begin{figure}[htbp]
\begin{center}
\includegraphics[width=\textwidth]{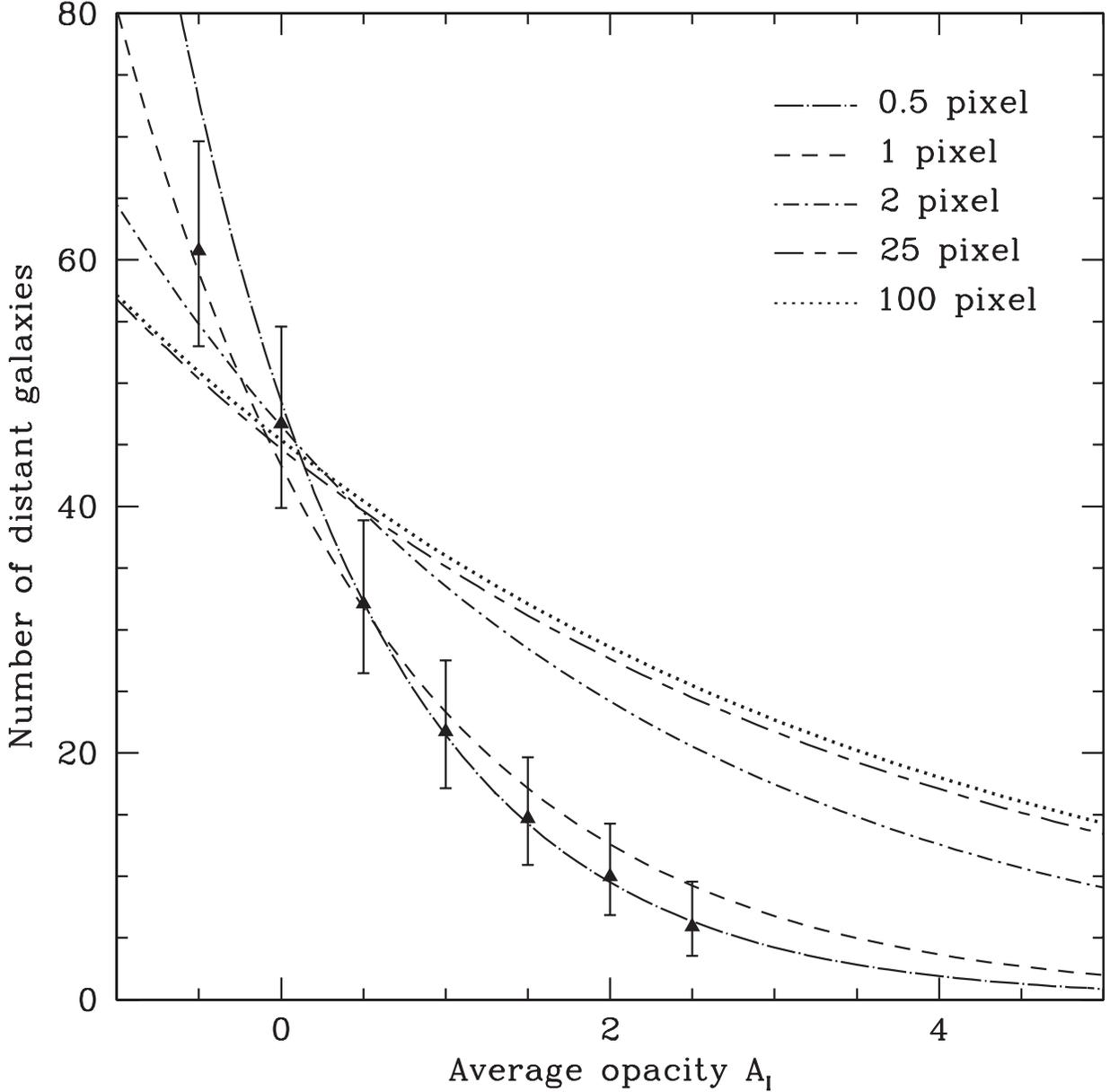}
\caption{Different synthetic fields with different dust disk models from \cite{Holwerda05}. A set of synthetic fields is made using dimmed HDF images. This dimming can be a smooth uniform gray screen (triangles), or some distribution of opaque clouds (curves). The relation between the number of distant galaxies from the HDF that can still be retrieved and the average opacity of the dimming depends on the assumed model. Large, resolved clouds block fewer background objects given the same filling factor. As a result, one a higher optical depth is implied by the intersection of the curve and number of observed background galaxies. The distance to NGC1365 is 18 Mpc, so a pixel of $0\farcs05$ is equivalent to 4 pc in linear size. The scales in the above simulations correspond, therefore, to clouds with cross-sections with a radius of 2, 4, 8, 100 and 400 pc, respectively  }
\label{fig:sfmsize}
\end{center}
\end{figure}

\begin{figure}[htbp]
\begin{center}
\includegraphics[width=\textwidth]{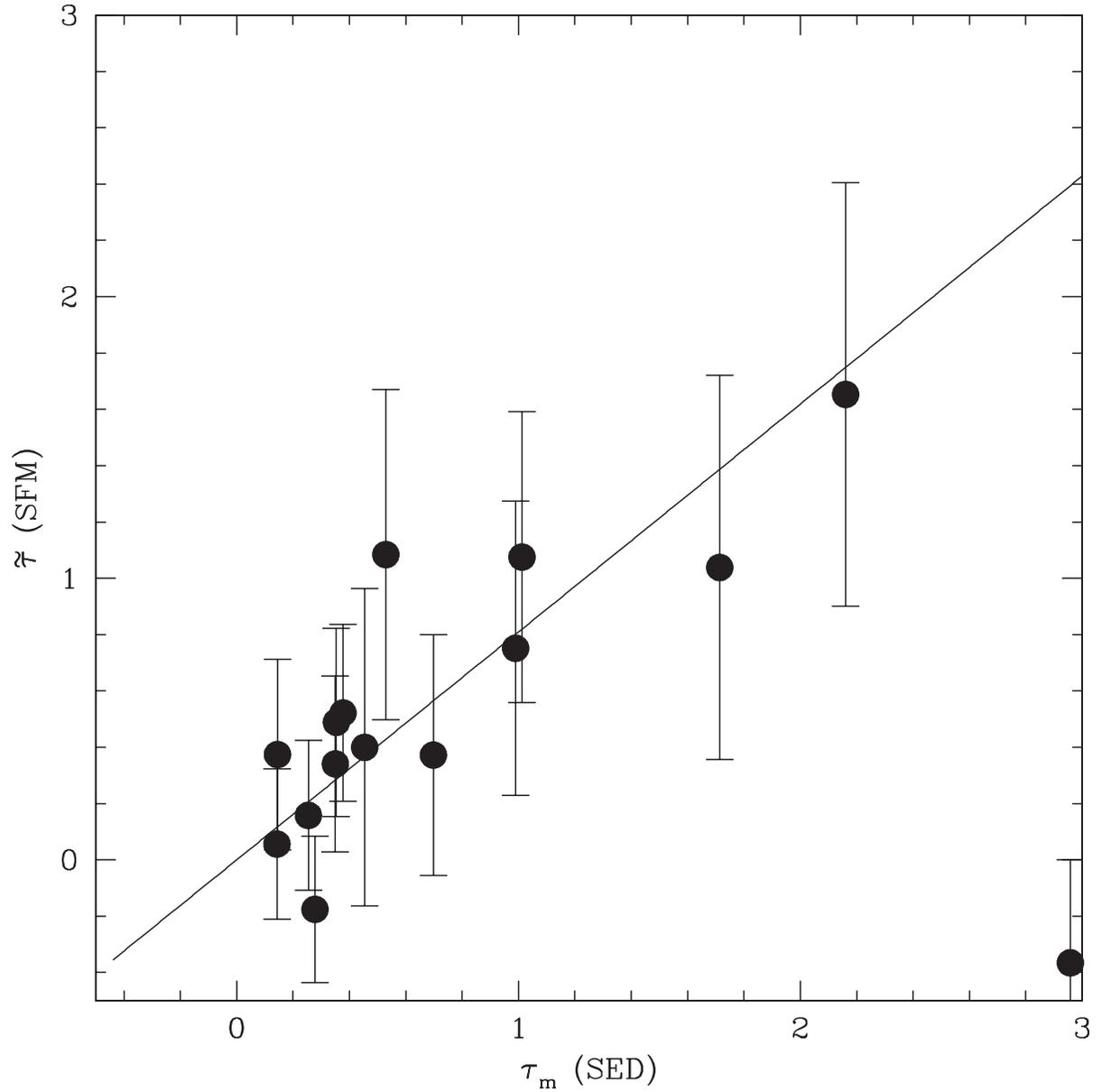}
\caption{The mean and apparent optical depth in the $I$-band, from SED ($\tau_m$) and SFM ($\tilde{\tau}$). The best fit with equation \ref{eq:rat} to the ratios of these values is also shown ($\tau_c = 0.4$). The average value for number of clouds along the line-of-sight, $n$, is then 2.6. If the negative SFM values are included in the fit, the cloud optical depth rises to $\tau_c = 0.6$ and the average $n$ becomes 1.9.}
\label{fig:fit}
\end{center}
\end{figure}

\begin{figure}[htbp]
\begin{center}
\includegraphics[width=\textwidth]{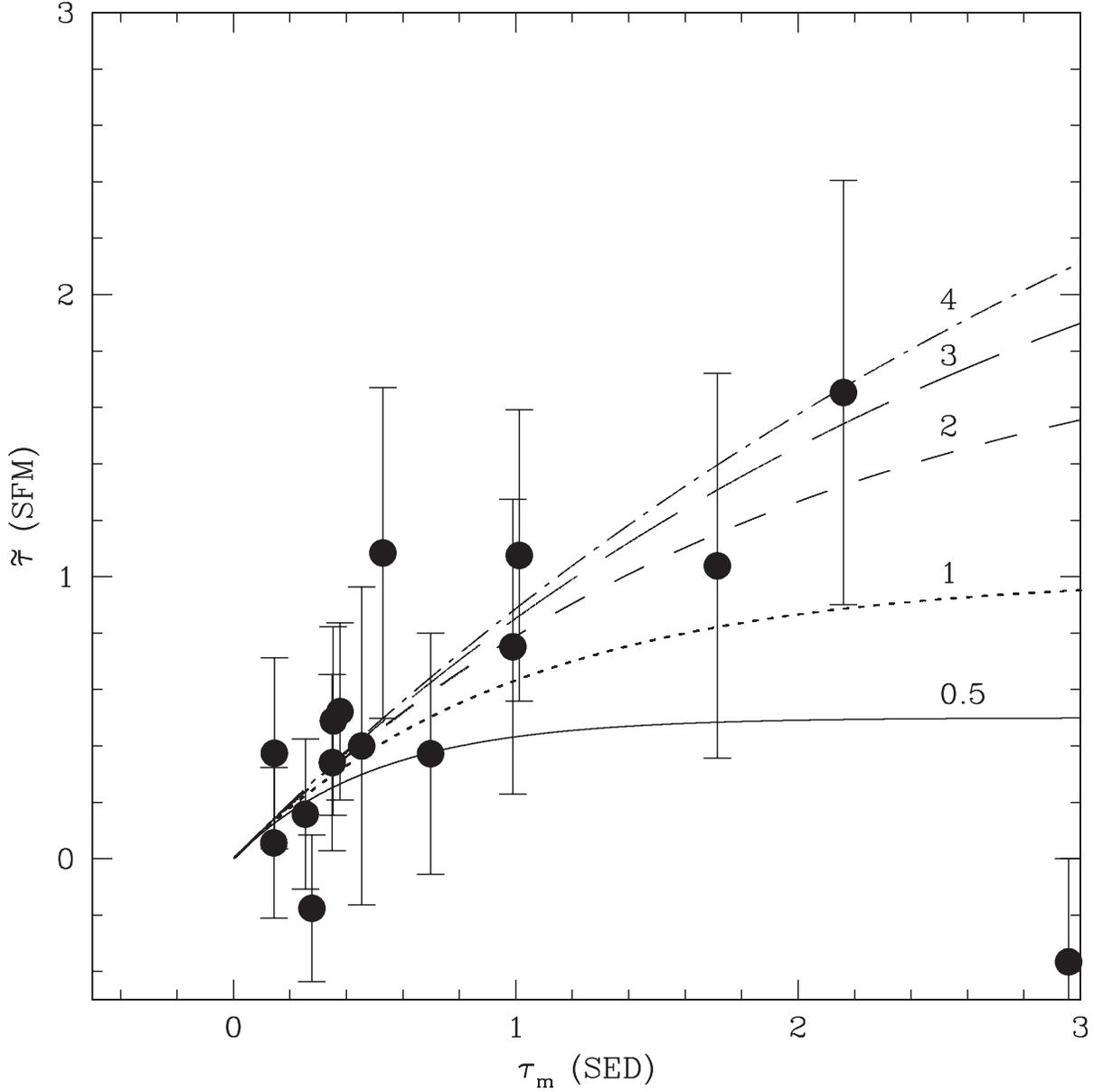}
\caption{Model values of $\tau_m$ and $\tilde{\tau}$, from equations \ref{eq:tm} and \ref{eq:tap}, for fixed values of the average number of clouds along the line-of-sight (n = 0.5, 1, 2, 3, 4). The cloud optical depth $\tau_c$ is left to vary proportionally to $\tau_m$ ($\tau_m = n \times \tau_c$, equation \ref{eq:tm}). Hence maximum cloud optical depths are 6, 3, 1.5, 1 and 0.75 respectively. If optically thin clouds are made up of optically thin clouds, it seems necessary that more than one cloud will lie along the line-of-sight.}
\label{fig:n_val}
\end{center}
\end{figure}

To constrain models of the spiral disk's energy budget with typical values for the size of dusty clouds 
in the ISM, we compare two techniques to extract the average and apparent optical 
depths of a section of spiral disk. From the comparison between SFM and SED results, we conclude the following:

\begin{itemize}
\item[1.] The SFM's calibration alone implies projected cloud scales predominantly unresolved with HST (of the order of 10 pc in NGC 1365, see \S \ref{sec:csize} and Figure \ref{fig:sfmsize}). 
\item[2.] The dust responsible for the disk's opacity is predominantly cold (T $<$ 25 K, Table \ref{table:model}).
\item[3.] The average and apparent optical depths of these disk sections, measured from SED and SFM respectively, generally agree (Figure \ref{fig:fit} and \ref{fig:n_val}). This implies generally optically thin clouds ($\tau_c <1$, \S \ref{ssec:tc}).
\item[4.] The fit to the ratio between apparent and average optical depth measurements, ${\tilde{\tau} / \tau_m}$, indicates a cloud optical depth, $\tau_c$, of 0.4, more than a single cloud along the line-of-sight, and a cloud size of $\sim$ 60 pc. 
If several partially overlapping clouds are seen in projection through the disk, the disk's opacity will change over smaller scales, consistent with conclusion 1.
\item[6.] Optically thick disks appear to have more than a single cloud along the line-of-sight (Figure \ref{fig:n_val}) and optically thin disks may have several clouds as well.
\end{itemize}

Future work using counts of distant galaxies through a foreground disk could be 
used to find cold dust structures at larger galactic radii, provided a sufficiently 
large solid angle has been imaged with HST/ACS's superb resolution.\footnote{See 
\cite{Holwerda05e} for selection criteria of suitable data.} Notably, the ACS 
data on M51, M81  and M101 are very promising for such an analysis.\footnote{The closer disks are slightly less well suited for the SFM because their stellar disk is resolved, but the loss in accuracy is offset by the larger available solid angle.}
{\it Spitzer} observations of these nearby disks are also available, making a similar comparison between SED and apparent optical depth possible for portions of these disks. The typical cloud scale for spiral arms or disk sections or as a function of galactic radius could then be found.
The SCUBA-2 instrument has recently been installed on the James Clerk Maxwell Telescope. 
A project with SCUBA-2 to map the SINGS galaxies in two sub-mm bands will improve 
future SED modelling of these spiral disks significantly over the SED models presented 
here. 

\acknowledgements

This work is based in part on archival data obtained with the {\it Spitzer} 
Space Telescope, which is operated by JPL, CalTech, under a contract 
with NASA. This work is also based on observations with the NASA/ESA Hubble Space 
Telescope, obtained at the STScI, which is operated by the Association 
of Universities for Research in Astronomy (AURA), Inc., under NASA 
contract NAS5-26555. 

The authors would like to thank T. Jarrett, for making his aperture corrections 
of extended sources available to us at an early stage, and Erik Rosolowsky, 
for useful discussions on Local Group cloud sizes. We would like to thank Maarten 
Baes for discussion of the motivation. We would also like to 
thank George Bendo, Erik Hollenback and Kristen Keener for their comments 
on earlier drafts of this paper.


%
\begin{deluxetable}{l l l l}
\tablewidth{0pt.}
\tablecaption{SINGS data description. IRAC and MIPS pixel scales, PSF FWHM, and aperture corrections (the factor by which the fluxes are multiplied). \label{table:data}}
\tablehead{
\colhead{Instrument} 	& \colhead{pixel}	& \colhead{PSF} 	& \colhead{aperture}\\
\colhead{(band)}		& \colhead{scale}	& \colhead{(FWHM)} 	& \colhead{correction}\\
	& \colhead{\tablenotemark{1}}	& \colhead{\tablenotemark{2}}	& \colhead{\tablenotemark{3}}}
\startdata
IRAC (3.6)  	 	& $0\farcs75$	& $2\farcs5$ 	& 0.9\\
IRAC (4.5)  	 	& $0\farcs75$	& $2\farcs5$ 	& 0.9\\
IRAC (5.8) 	 	& $0\farcs75$	& $2\farcs5$ 	& 0.7\\
IRAC (8.0)  	 	& $0\farcs75$	& $2\farcs5$ 	& 0.75\\

MIPS (24)		 	& $1\farcs5$	& $6\farcs0$ 	& 1.16\\
MIPS (70) 	 	& $4\farcs5$	& $18\farcs0$ 	& 1.2\\
MIPS (160) 	 	& $9\farcs0$	& $40\farcs0$ 	& 1.4\\
\enddata
\tablenotetext{1}{Pixel scales were set by the SINGS team.}
\tablenotetext{2}{FWHM values for the IRAC are conservative estimates. Actual FWHM values are better than $2\farcs5$. }
\tablenotetext{3}{The IRAC values are the initial results from T. Jarrett ({\it private communication}), but they do not differ substantially from the final results.}
\end{deluxetable}

%
\begin{deluxetable}{l l l l l l l l}
\tablewidth{0pt.}
\tabletypesize{\scriptsize}
\tablecaption{{\it Spitzer} channel surface brightness in the WFPC2 aperture (3 WFs $\times 1.3\arcmin \times 1.3\arcmin$).\label{table:flux}}
\tablehead{
\colhead{Galaxy}		& \colhead{3.6 $\rm \mu m$}		& \colhead{4.5 $\rm \mu m$} & \colhead{5.8 $\rm \mu m$} & \colhead{ 8 $\rm \mu m$} & \colhead{24 $\rm \mu m$} & \colhead{70 $\rm \mu m$} & \colhead{160 $\rm \mu m$} \\
	& \colhead{MJy/sr} 	& \colhead{MJy/sr} 	& \colhead{MJy/sr} 	& \colhead{MJy/sr} 	& \colhead{MJy/sr} 	& \colhead{MJy/sr} 	& \colhead{MJy/sr} 	\\
}
\startdata
NGC 925		& 0.17 $\pm$ 0.04   & 0.19 $\pm$ 0.08   & 0.12 $\pm$ 0.05   & 3.79 $\pm$ 2.65   & 0.33 $\pm$ 0.06   & 4.76 $\pm$ 0.44   & 21.92 $\pm$ 0.91  \\
NGC 2841	& 0.39 $\pm$ 0.04   & 0.33 $\pm$ 0.09   & 0.88 $\pm$ 0.59   & 3.42 $\pm$ 2.43   & 0.46 $\pm$ 0.04   & 4.54 $\pm$ 0.34   & 34.27 $\pm$ 0.58  \\ 
NGC 3031	& 0.56 $\pm$ 0.03   & 0.39 $\pm$ 0.06   & 0.25 $\pm$ 0.07   & 2.07 $\pm$ 1.41   & 0.23 $\pm$ 0.04   & 3.74 $\pm$ 0.37   & 23.73 $\pm$ 1.38  \\ 
NGC 3198	& 0.13 $\pm$ 0.02   & 0.12 $\pm$ 0.04   & 0.11 $\pm$ 0.04   & 2.28 $\pm$ 1.20   & 0.32 $\pm$ 0.05   & 3.48 $\pm$ 0.36   & 20.09 $\pm$ 0.46  \\ 
NGC 3351	& 0.55 $\pm$ 0.05   & 0.45 $\pm$ 0.16   & 0.33 $\pm$ 0.07   & 5.81 $\pm$ 4.14   & 1.11 $\pm$ 0.06   & 12.54 $\pm$ 0.54  & 43.55 $\pm$ 0.68  \\ 
NGC 3621-1	& 0.24 $\pm$ 0.01   & 0.12 $\pm$ 0.01   & 0.40 $\pm$ 0.05   & 1.27 $\pm$ 0.37   & 0.70 $\pm$ 0.05   & 9.68 $\pm$ 0.46   & 41.38 $\pm$ 0.87  \\ 
NGC 3621-2	& 0.15 $\pm$ 0.01   & 0.17 $\pm$ 0.01   & 0.24 $\pm$ 0.05   & 1.08 $\pm$ 0.37   & 0.66 $\pm$ 0.05   & 11.04 $\pm$ 0.46  & 36.81 $\pm$ 0.87  \\
NGC 3627	& 0.97 $\pm$ 0.04   & 0.68 $\pm$ 0.06   & 1.20 $\pm$ 0.06   & 6.72 $\pm$ 2.89   & 5.80 $\pm$ 0.06   & 49.58 $\pm$ 0.61  & 152.63 $\pm$ 0.95 \\ 
NGC 4321	& 0.55 $\pm$ 0.02   & 0.39 $\pm$ 0.04   & 0.64 $\pm$ 0.06   & 4.45 $\pm$ 2.23   & 2.62 $\pm$ 0.05   & 25.12 $\pm$ 0.46  & 100.53 $\pm$ 0.75 \\ 
NGC 4536	& 0.11 $\pm$ 0.04   & 0.15 $\pm$ 0.07   & 0.72 $\pm$ 3.45   & 3.64 $\pm$ 2.89   & 0.19 $\pm$ 0.05   & 2.35 $\pm$ 0.52   & 10.97 $\pm$ 0.60  \\ 
NGC 4559	& 0.05 $\pm$ 0.03   & 0.04 $\pm$ 0.01   & 0.06 $\pm$ 0.05   & 1.51 $\pm$ 1.05   & 0.10 $\pm$ 0.05   & 2.65 $\pm$ 0.38   & 11.80 $\pm$ 0.53  \\ 
NGC 4725	& 0.11 $\pm$ 0.02   & 0.09 $\pm$ 0.05   & 0.28 $\pm$ 0.20   & 1.76 $\pm$ 1.29   & 0.16 $\pm$ 0.05   & 1.59 $\pm$ 0.35   & 12.50 $\pm$ 0.49  \\
NGC 5194-1	& 0.52 $\pm$ 0.01   & 0.32 $\pm$ 0.01   & 0.80 $\pm$ 0.05   & 2.61 $\pm$ 0.22   & 2.64 $\pm$ 0.04   & 28.61 $\pm$ 0.38  & 124.68 $\pm$ 1.36 \\ 
NGC 5194-2	& 0.56 $\pm$ 0.01   & 0.34 $\pm$ 0.01   & 0.88 $\pm$ 0.05   & 2.76 $\pm$ 0.22   & 3.26 $\pm$ 0.04   & 36.35 $\pm$ 0.38  & 144.30 $\pm$ 1.36 \\ 
NGC 6946	& 0.43 $\pm$ 0.03   & 0.27 $\pm$ 0.01   & 0.71 $\pm$ 0.09   & 2.31 $\pm$ 0.48   & 2.92 $\pm$ 0.07   & 28.88 $\pm$ 0.53  & 80.83 $\pm$ 2.13  \\ 
NGC 7331	& 0.14 $\pm$ 0.01   & 0.10 $\pm$ 0.01   & 0.10 $\pm$ 0.05   & 1.39 $\pm$ 0.88   & 0.23 $\pm$ 0.04   & 3.49 $\pm$ 0.67   & 17.05 $\pm$ 0.79  \\
\enddata
\end{deluxetable}

%
\begin{deluxetable}{l c c c c c c c c c c}
\tablewidth{0pt.}
\tabletypesize{\scriptsize}

\tablecaption{Model output: \label{table:model} stellar and dust surface brightness, dust surface density, dust mean temperature \citep[$T_2$ in][]{Draine07a}, fit quality, the parameters of stellar irradiation --minimum and maximum of the distribution, mean irradiative field, fraction of dust exposed to more than $U_{min}$. Models can be found at http://www.astro.princeton.edu/$\sim$draine/dust/irem.html }
\tablehead{
\colhead{Galaxy} 	& \colhead{$L_{star}/area$}	   & \colhead{$L_{dust}/area$} 		& \colhead{$M_{dust}/area$} & \colhead{T} &   \colhead{$\chi^2$}& \colhead{$U_{min}$} & \colhead{$U_{max}$} & \colhead{$\bar{U}$} & \colhead{$\gamma$} & \colhead{model} \\
\colhead{name}         & \colhead{($L_{\odot}/kpc^2$)} & \colhead{($L_{\odot}/kpc^2$)}    & \colhead{($ M_{\odot}/kpc^2$)}   & \colhead{(K)} & \colhead{} & \colhead{}& \colhead{}& \colhead{}& \colhead{} & \colhead{name}\\ 
\colhead{}         & \colhead{$\times 10^8$} & \colhead{$\times 10^8$}    & \colhead{$\times 10^6$}   & \colhead{} & \colhead{} & \colhead{} & \colhead{$\times 10^6$} & \colhead{}  & \colhead{$\times 10^{-3}$} & 
}
\startdata
NGC 0925	& 0.8	& 0.2	& 0.2	& 15.7	& 2.8 	& 0.70 & 10  & 0.8  & 5.8	& U0.70\_1e7\_MW3.1\_20  \\
NGC 2841	& 1.8	& 0.3	& 0.5	& 14.6	& 4.15 	& 0.70 & 10  & 0.8  & 1.0	& U0.70\_1e7\_MW3.1\_20 \\
NGC 3031	& 2.7	& 0.2	& 0.3	& 14.6	& 2.32 	& 0.50 & 1.0 & 0.5  & 1.5	& U0.50\_1e6\_MW3.1\_30\\
NGC 3198	& 0.6	& 0.2	& 0.3	& 15.0	& 3.73 	& 0.50 & 10  & 0.6  & 7.5	& U0.50\_1e7\_MW3.1\_30 \\
NGC 3351	& 2.7	& 0.5	& 0.4	& 16.2	& 1.9 	& 0.80 & 1.0 & 1.0  & 1.6	& U0.80\_1e6\_MW3.1\_10\\
NGC 3621-1	& 0.7	& 0.5	& 0.4	& 16.2	& 2.45 	& 1.00 & 1.0 & 1.0  & 2.6	& U1.00\_1e6\_MW3.1\_60\\
NGC 3621-2	& 0.7	& 0.4	& 0.2	& 16.9	& 10.91	& 1.20 & 10  & 1.3  & 7.3	& U1.20\_1e7\_MW3.1\_30 \\
NGC 3627	& 3.7	& 2.1	& 1.1	& 17.0	& 2.11 	& 1.00 & 10  & 1.4  & 24.5	& U1.00\_1e7\_MW3.1\_30\\
NGC 4321	& 2.2	& 1.1	& 1.0	& 16.0	& 1.91 	& 0.70 & 10  & 0.9  & 15.6	& U0.70\_1e7\_MW3.1\_30 \\
NGC 4536	& 0.5	& 0.1	& 0.1	& 15.7	& 2.32 	& 0.80 & 10  & 0.8  & 2.7	& U0.80\_1e7\_MW3.1\_60 \\
NGC 4559	& 0.2	& 0.1	& 0.1	& 15.7	& 1.76 	& 0.80 & 1.0 & 0.8  & 0.0	& U0.80\_1e6\_MW3.1\_30  \\
NGC 4725	& 0.5	& 0.1	& 0.2	& 14.6	& 3.01 	& 0.50 & 1.0 & 0.5  & 0.0	& U0.50\_1e6\_MW3.1\_60  \\
NGC 5194-1	& 1.6	& 1.4	& 1.2	& 16.0	& 2.33 	& 0.80 & 1.0 & 0.9  & 9.9 	& U0.80\_1e6\_MW3.1\_50    \\
NGC 5194-2	& 1.8	& 1.7	& 1.4	& 16.0	& 1.83 	& 0.80 & 1.0 & 0.9  & 11.7 & U0.80\_1e6\_MW3.1\_40  \\
NGC 6946	& 1.4	& 1.2	& 0.5	& 17.8	& 0.65 	& 1.50 & 1.0 & 1.9  & 20.5	& U1.50\_1e6\_MW3.1\_50\\
NGC 7331	& 0.6	& 0.2	& 0.2	& 15.4	& 2.15 	& 0.70 & 10  & 0.7  & 4.5 	& U0.70\_1e7\_MW3.1\_30\\

\enddata
\end{deluxetable}

%
\begin{deluxetable}{l l l l l l l}
\tablewidth{0pt.}
\tablecaption{The apparent ($\tilde{\tau}$) and average ($\tau_m $) optical depths in $I$-band measured in the WFPC2 field (3 WFs $1\farcm3 \times 1\farcm3$), uncorrected and corrected for inclination.  \label{table:A}}
\tablehead{
\colhead{Galaxy}		& \colhead{$\tilde{\tau}$}	& \colhead{$\tilde{\tau}$}	& \colhead{$\tau_m $} &  \colhead{$\tau_m $}& \colhead{$\tilde{\tau}/\tau_m$}\\
					& \colhead{(SFM)} 		& \colhead{$\times cos(i)$}& \colhead{(SED)} & \colhead{$\times cos(i)$} & }
\startdata
NGC 0925	& $-0.4^{0.3}_{-0.3}$ & -0.2 & 0.6	& 0.3 & -0.63 \\
NGC 2841	& $0.7^{0.4}_{-0.4}$	& 0.4	& 1.4	& 0.7 & 0.53 \\
NGC 3031	& $0.8^{0.5}_{-0.6}$	& 0.4	& 0.9	& 0.5 & 0.88 \\
NGC 3198	& $0.7^{0.3}_{-0.3}$	& 0.3	& 0.7	& 0.4 & 0.97 \\
NGC 3351	& $1.1^{0.5}_{-0.5}$	& 1.1	& 1.0	& 1.0 & 1.06 \\
NGC 3621-1	& $2.0^{0.5}_{-0.6}$	& 1.1	& 1.0	& 0.5 & 2.05 \\
NGC 3621-2	& $1.0^{0.3}_{-0.3}$	& 0.5	& 0.7	& 0.4 & 1.38 \\
NGC 3627	& $1.9^{0.6}_{-0.7}$	& 1.0	& 3.1 	& 1.7 & 0.61 \\ 
NGC 4321	& $2.2^{0.7}_{-0.8}$	& 1.7	& 2.8 	& 2.2 & 0.76\\ 
NGC 4536	& $0.8^{0.3}_{-0.3}$	& 0.4	& 0.3	& 0.1 & 2.58 \\ 
NGC 4559	& $0.1^{0.3}_{-0.3}$	& 0.1	& 0.3	& 0.1 & 0.39 \\
NGC 4725	& $0.7^{0.3}_{-0.3}$	& 0.5	& 0.5	& 0.4 & 1.38 \\
NGC 5194-1	& $-0.4^{0.4}_{-0.4}$ & -0.4	& 3.3	& 3.0 & -0.12 \\
NGC 5194-1	& $1.3^{0.5}_{-0.6}$	& 1.2	& 3.9	& 3.5 & 0.34 \\
NGC 6946	& $1.0^{0.5}_{-0.5}$	& 0.8	& 1.3	& 1.0 & 0.76 \\
NGC 7331	& $0.3^{0.3}_{-0.3}$	& 0.2	& 0.5	& 0.3 & 0.62 \\
\enddata
\end{deluxetable}

\end{document}